\documentclass{moriond}

\def\Journal#1#2#3#4{{#1} {\bf #2}, #3 (#4)}

\def\NIMA{{\em Nucl. Instrum. Methods} A}

\def\PLB{{\em Phys. Lett.} B}

\def\JHEP{\em JHEP}
\def\EPJC{{\em Eur. Phys. J.} C}

\def\be{\begin{equation}}
\def\ee{\end{equation}}
\def\bea{\begin{eqnarray}}
\def\eea{\end{eqnarray}}

\def\tauhad{\ensuremath{\tau_\mathrm{had}}}

\newcommand{\Photo}{\includegraphics[height=35mm]{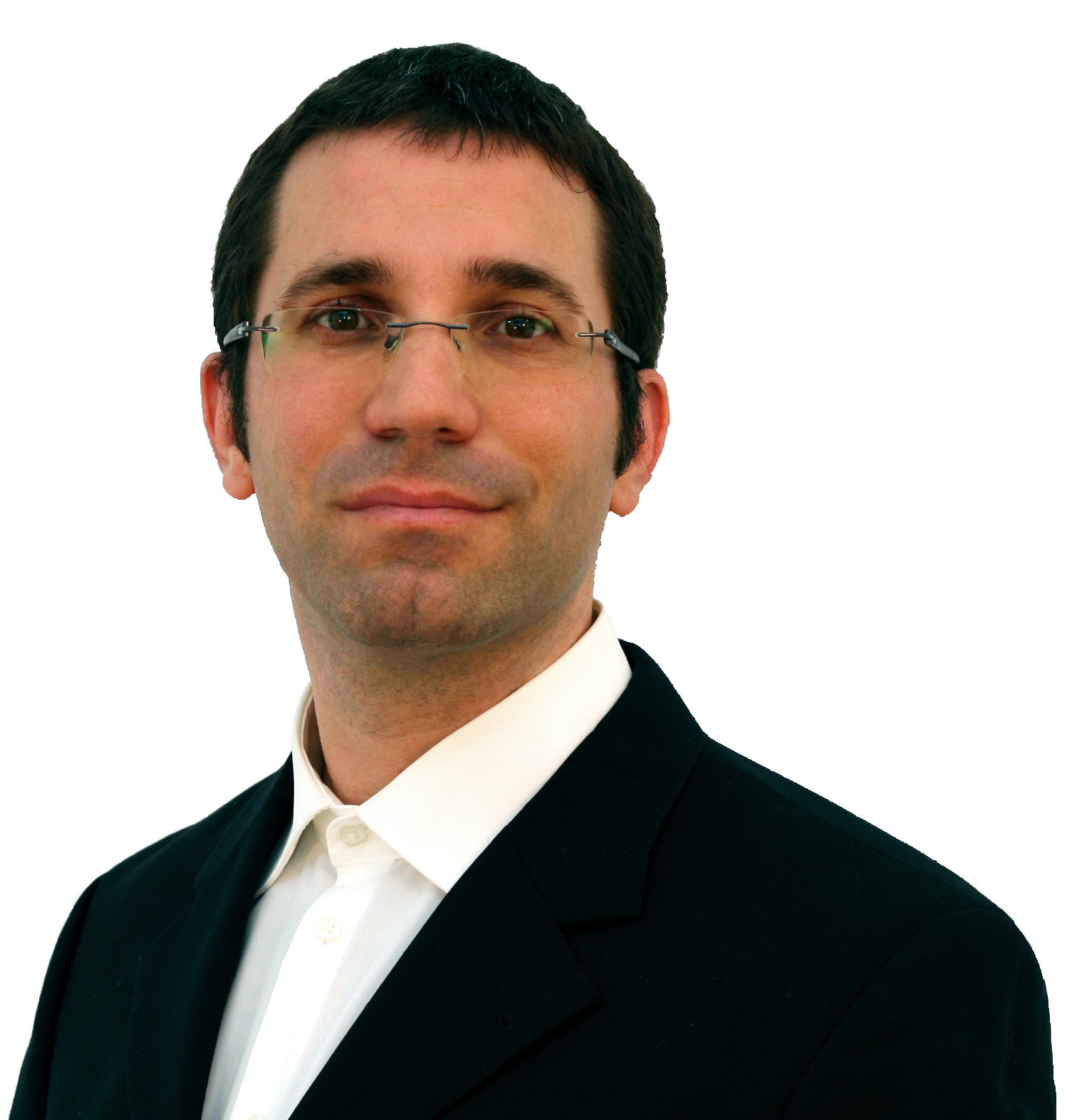}}

\begin{document}
\vspace*{4cm}
\title{CMS Higgs physics results}

\author{ M. Flechl \emph{on behalf of the CMS collaboration} }

\address{Institute of High-Energy Physics,
Austrian Academy of Sciences,\\
Nikolsdorfer Gasse 18,
1220 Vienna, Austria}  

\maketitle\abstracts{
Recent results of searches for Higgs bosons by the CMS collaboration are presented. These consist of searches for rare 
Higgs boson decays, searches for additional neutral Higgs bosons, and searches for charged Higgs bosons.
}

\section{Introduction}
A Higgs boson with a mass of about 125 GeV has been discovered by the ATLAS and CMS collaborations in 2012\cite{atlash,cmsh,cmsh2}. 
Since then, the amount of analyzed data has significantly increased and the analysis methods have been adapted, 
either for measuring Higgs boson properties or for searching for additional Higgs bosons considering the information 
we have about the discovered Higgs boson, H-125, in particular its mass. So far, all measurements are in agreement with the 
standard model (SM) hypothesis; however, the sensitivity to many beyond-the-SM (BSM) scenarios is still very patchy and the 
analysis of the full LHC Run 2 data as well as of following run periods will be needed to test them. 
In addition to collecting more data, new final states are being probed and provide additional handles on BSM physics. 
Those include adding measurements of rare H-125 boson decay modes and adding searches for additional neutral and 
charged Higgs bosons in decay modes not investigated previously.

This articles summarizes the results of CMS\cite{cms} searches for rare Higgs boson decay modes, including Higgs boson 
decays to invisible particles, Higgs boson decays to mesons and to light pseudoscalars; searches for additional 
neutral Higgs bosons (scalar H or pseudoscalar A) in decays to a top quark pair or to ZH, and charged Higgs boson decays to $\tau\nu$, 
tb, and AW.

\section{Rare Higgs decays}
\subsection{Higgs boson production with a top quark pair and decay to invisible particles}
\begin{figure}[h]
\begin{minipage}{0.30\linewidth}
\centerline{\includegraphics[width=1.0\linewidth]{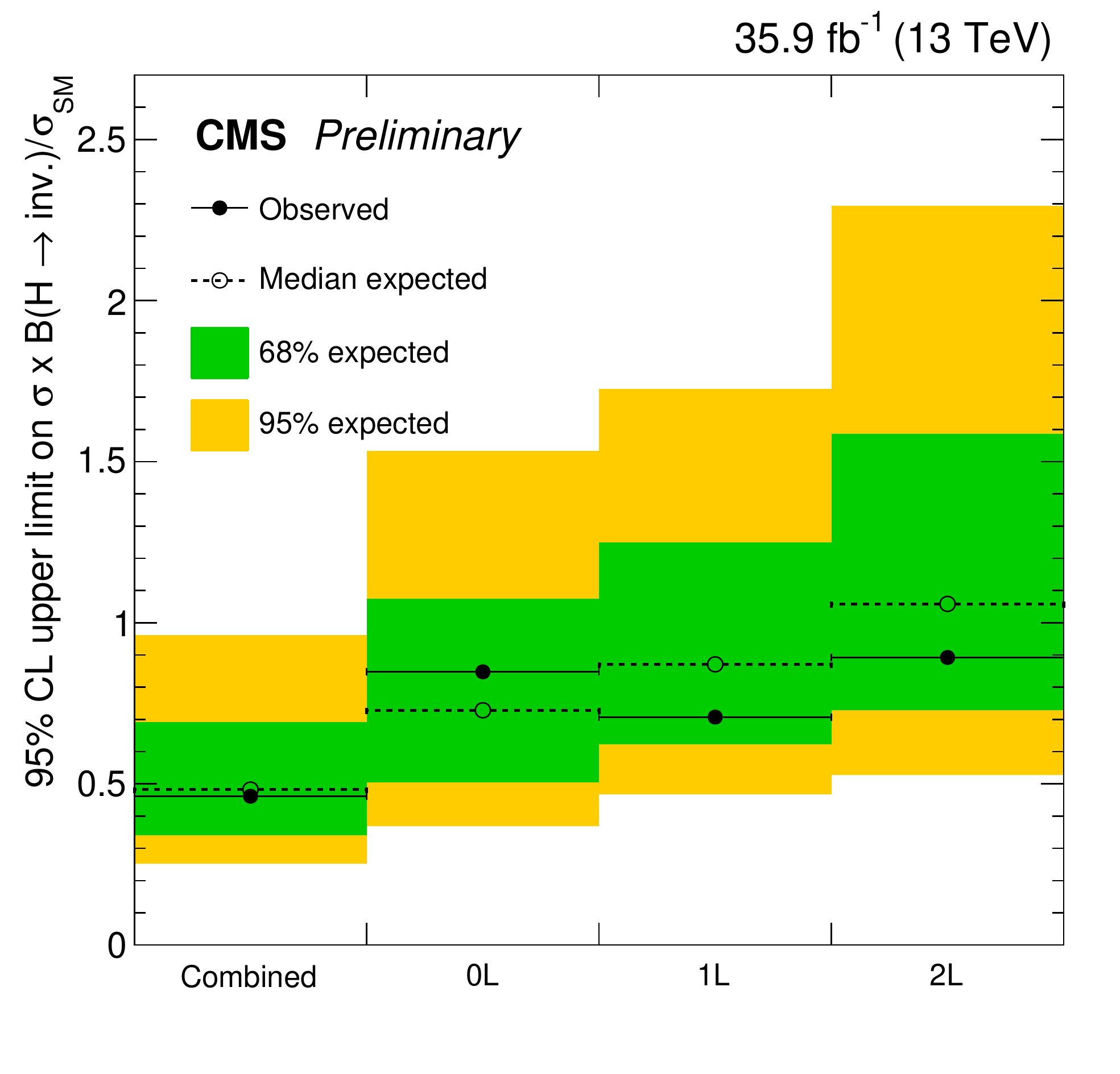}}
\end{minipage}
\hfill
\begin{minipage}{0.30\linewidth}
\centerline{\includegraphics[width=1.0\linewidth]{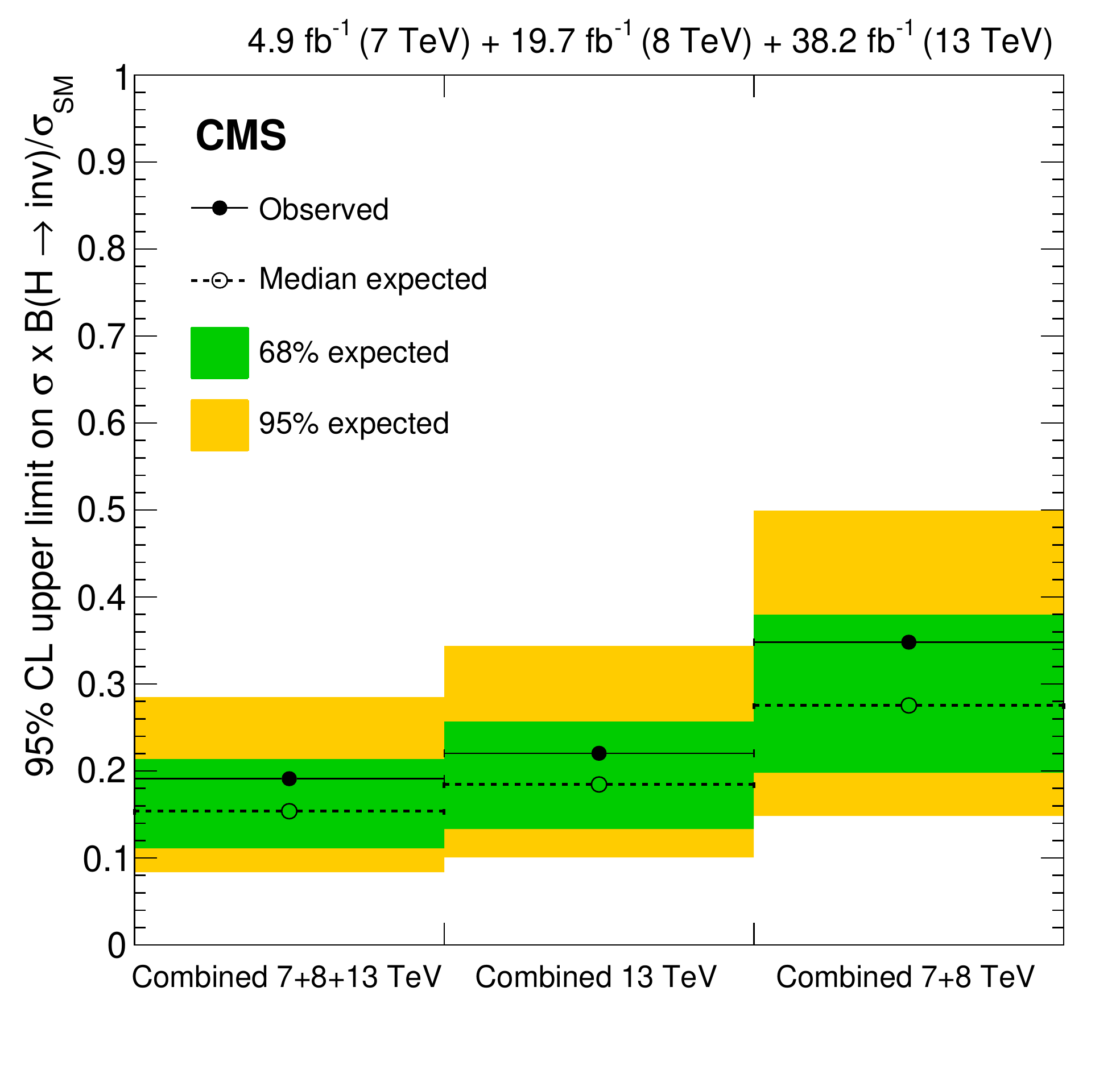}}
\end{minipage}
\hfill
\begin{minipage}{0.38\linewidth}
\centerline{\includegraphics[width=1.0\linewidth]{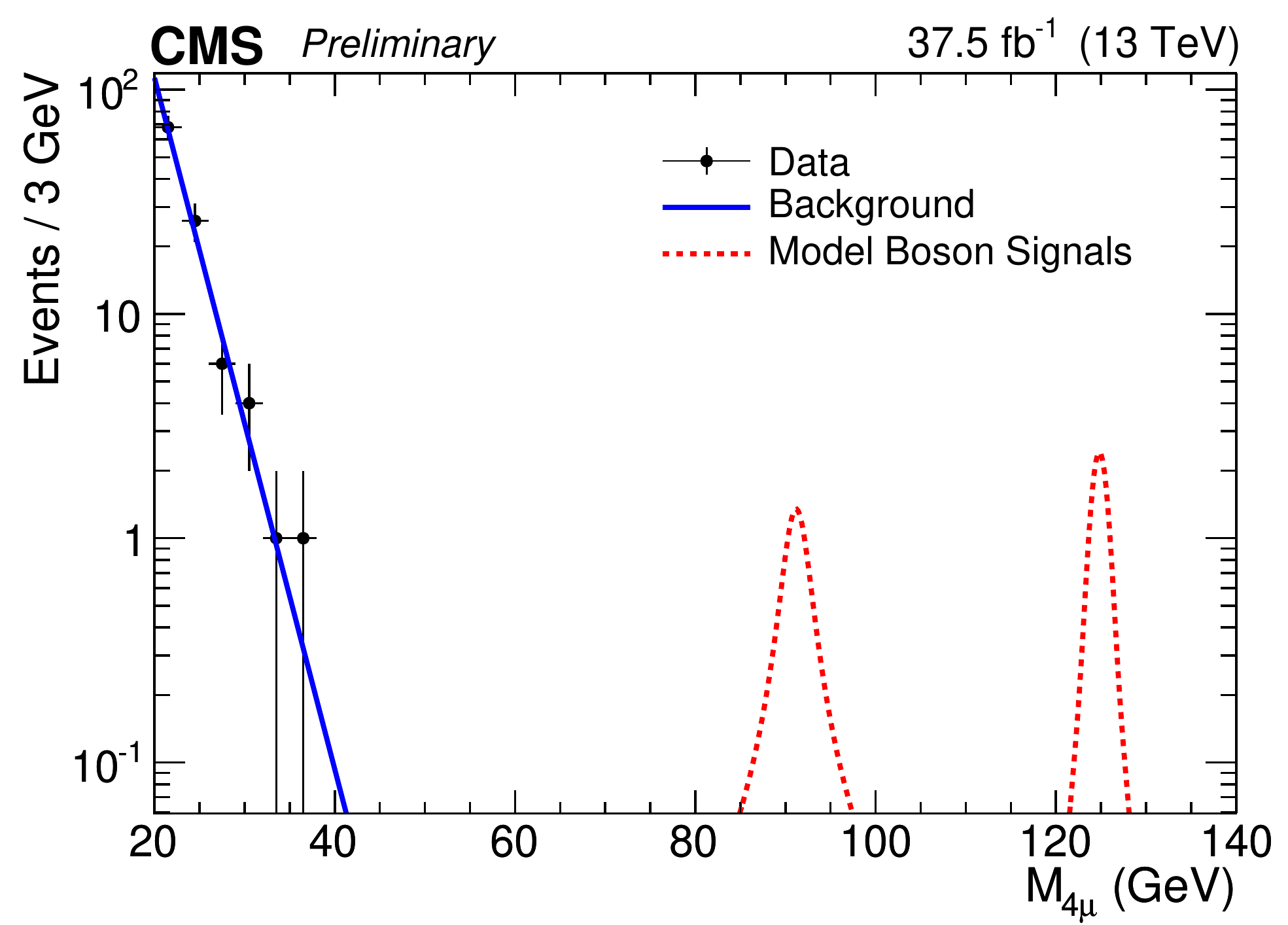}}
\end{minipage}
\caption[]{Limit on the branching ratio H$\to$inv measured in a ttH topology\cite{tth_inv} (left), limit on the branching ratio H$\to$inv combining measurements 
in gF, VBF and VH topologies\cite{comb_inv} (center) and four-muon-mass distribution in the search for Higgs boson decays to $\Upsilon\Upsilon$\cite{hmesons} (right).}
\label{fig1}
\end{figure}

First bounds on Higgs boson decays to invisible particles in a topology with two top quarks (ttH) 
are obtained by reinterpreting searches for scalar top quarks in the 0-lepton (0L), 1-lepton (1L) and 
2-lepton (2L) channels. The final state in all cases consists of a varying amount of leptons, jets and 
missing transverse momentum. A combined limit of B(H$\to$inv)$<$0.46 is measured (expected: 0.48)\cite{tth_inv}, 
see Figure~\ref{fig1}. This is comparable to the sensitivity in other topologies.

\subsection{Combined results for Higgs boson decays to invisible particles}
The current best limit on invisible Higgs boson decays is obtained by a combination of analyses using CMS data taken during LHC Run 1 as well as in 2015 and 2016\cite{comb_inv}. 
These analyses investigate the gluon fusion (gF), vector boson fusion (VBF) and V-associated production (VH). The limit is measured as B(H$\to$inv)$<$0.19 (expected: 0.15) 
and is shown in Figure~\ref{fig1}. Adding the LHC Run 1 data improves the sensitivity by about 5 per cent points. The results are also interpreted in the context of dark-portal models and 
are competitive to direct dark matter searches for dark matter masses below about 10 GeV.

\subsection{Higgs boson decays to mesons}
Searches for Higgs boson decays to pairs of J$/\Psi$ or $\Upsilon$ mesons are almost background-free and hence their sensitivity will increase quickly with additional data. 
Since the SM prediction for the branching ratio to these mesons is very small (about $10^{-9} - 10^{-10}$), any beyond-the-SM enhancement like new amplitudes would be 
immediately discernible if it is sufficiently large. For the analyses\cite{hmesons}, dedicated 2- and 3-muon triggers with meson mass windows for the muon systems are used. 
Muons with transverse momenta above 3 GeV are considered. The four-muon mass distribution for the $\Upsilon\Upsilon$ analysis is shown in Figure~\ref{fig1}. The 
measured limits are B$(\mathrm{H}\to \mathrm{J}\Psi \mathrm{J}\Psi)<1.8 \times 10^{-3}$ (expected: $1.8 \times 10^{-3}$) and B$(\mathrm{H}\to \Upsilon \Upsilon)<1.4 \times 10^{-3}$ 
(expected: $1.4 \times 10^{-3}$).

\subsection{Higgs boson decays to light pseudoscalars}
\begin{figure}
\begin{minipage}{0.31\linewidth}
\centerline{\includegraphics[width=1.00\linewidth]{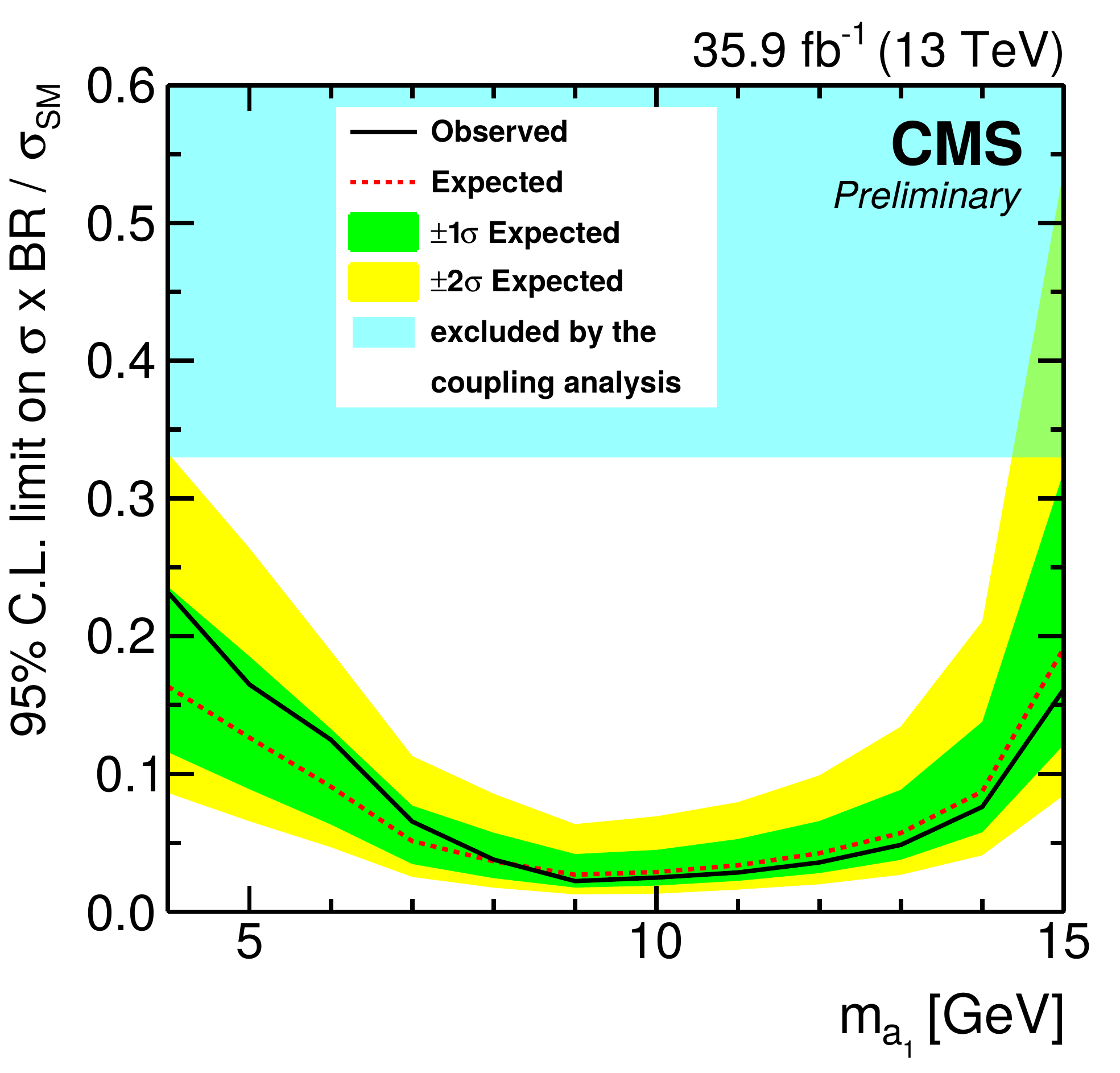}}
\end{minipage}
\hfill
\begin{minipage}{0.27\linewidth}
\centerline{\includegraphics[width=1.00\linewidth]{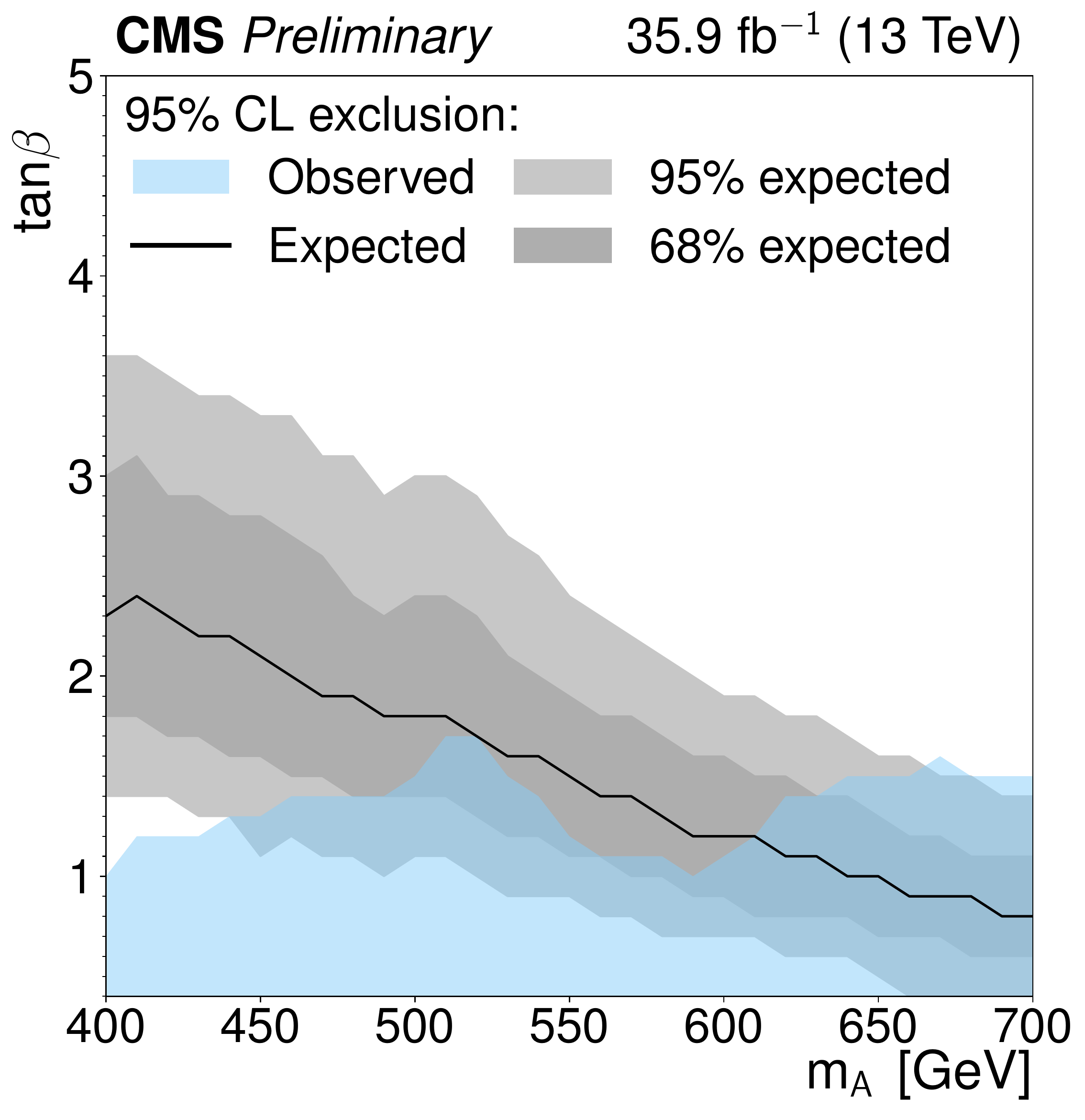}}
\end{minipage}
\hfill
\begin{minipage}{0.40\linewidth}
\centerline{\includegraphics[width=1.00\linewidth]{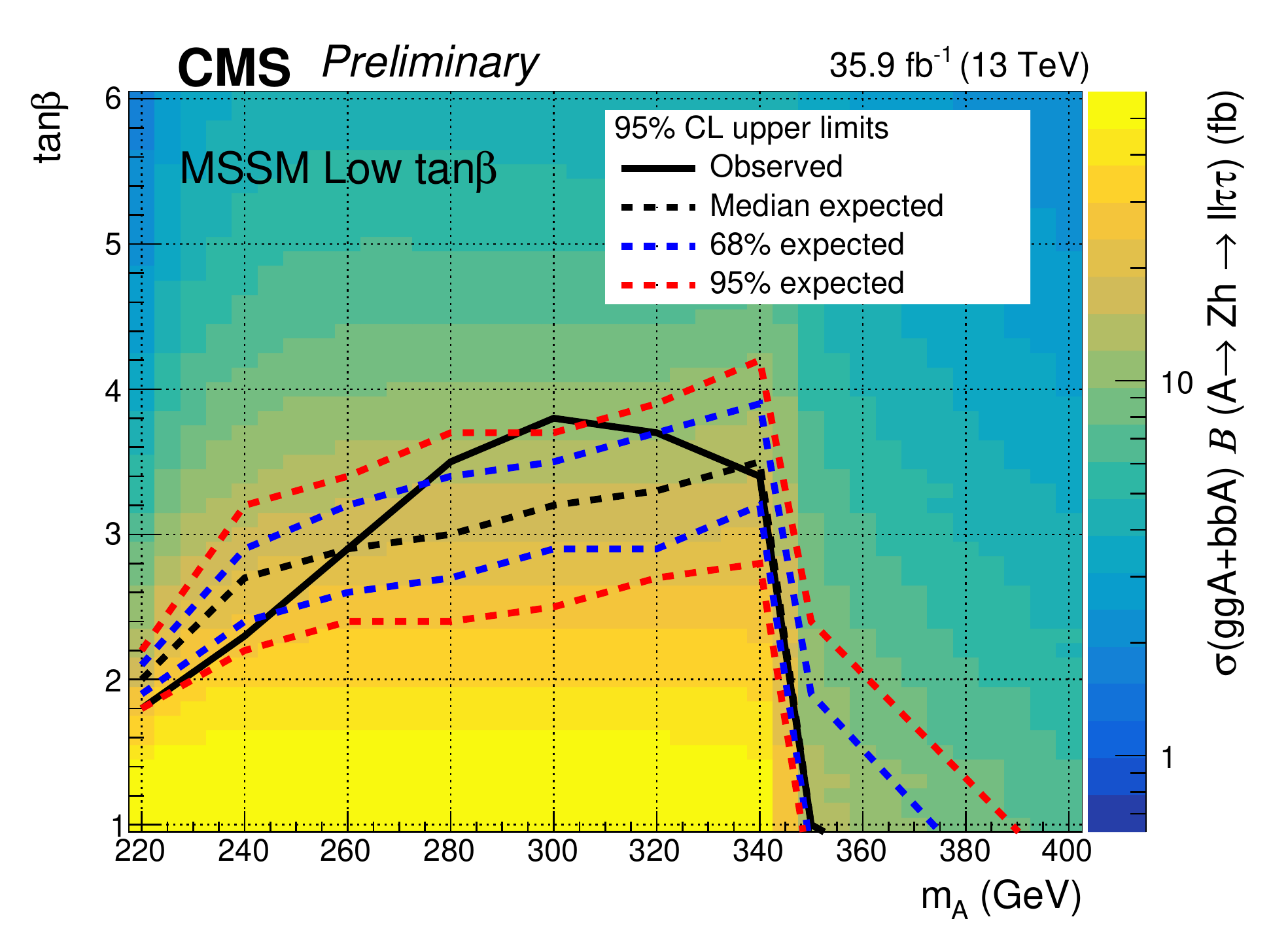}}
\end{minipage}
\caption[]{Limit on Higgs boson decays to a pair of pseudoscalars\cite{haa} (left), interpretation of the search for A/H decays to a top quark pair\cite{httbar} 
in the hMSSM scenario (center) and exclusion region of the A$\to$Zh analysis in the MSSM-low-$\tan\beta$ scenario\cite{azh} (right).}
\label{fig2}
\end{figure}

Searches for H-125 decays to a pair of light pseudoscalars, a$_1$, are mainly motivated by scenarios with a 2-Higgs-doublet model (2HDM) and an additional 
singlet, for example the next-to-minimal supersymmetric extension of the SM (NMSSM). The analysis presented here\cite{haa} studies the a$_1$ mass range from 4 GeV to 15 GeV 
and a$_1$ decays to a pair of tau leptons or muons which can have a sizable branching ratio in this mass range. One of the challenges is that such light 
a$_1$ mesons would be highly boosted and hence its decay products likely to geometrically overlap. The resulting limit as a function of the a$_1$ mass is shown 
in Figure~\ref{fig2} and improves previous CMS limits significantly beyond pure luminosity scaling, e.g. for $m_\mathrm{A} = 8$ GeV the branching ratio limits improves from 
about 25\% to about 3\%.

\section{Heavy neutral Higgs boson searches}
\subsection{Heavy Higgs boson decays to a top quark pair}
In the minimal supersymmetric extension of the SM (MSSM), for a low value of $\tan \beta$ (the ratio of the Higgs vacuum expectation values) and a heavy Higgs 
boson A or H with a mass of 
at least twice the top quark mass, the search for Higgs boson decays to a pair of top quarks might be the only way to see additional Higgs bosons at the LHC 
since the branching ratio can approach unity. The main challenge for this search is the interference of the signal with SM top quark pair production which leads to 
a peak-dip structure of the generated Higgs boson mass spectrum (instead of just a peak at the Higgs boson mass) and requires a non-linear signal model for the interpretation. 
The analysis sees a slight excess/deficit combination with a local significance of $3.8\sigma$ which would be compatible with a Higgs boson of a mass of 400 GeV\cite{httbar}. 
However, the global significance is only $2\sigma$. This also manifests itself as a difference between expected and observed exclusion region in the interpretation 
of the result in the hMSSM\cite{hmssm} scenario, see Fig.~\ref{fig2}.

\subsection{Heavy Higgs boson decays to Zh}
If $\tan \beta$ is small but the pseudoscalar A is too light to decay to a pair of top quarks then the decay mode A$\to$Zh could be dominant with a branching ratio close to unity. 
The analysis in the $\ell\ell\tau\tau$ final state targets primarily gluon fusion production and the 125-GeV Higgs boson in the final state\cite{azh}. The A boson mass 
is estimated using a matrix-element-based estimator (SVFit\cite{svfit}) which also exploits the 125-GeV constraint. The observed distributions agree well with the 
SM-only hypothesis and MSSM limits are set in the expected low $\tan\beta$ mass region from 220 GeV to 350 GeV, see Figure~\ref{fig2}.

\section{Charged Higgs boson searches}

\subsection{Charged Higgs boson decays to $\tau\nu$}
\begin{figure}
\begin{minipage}{0.32\linewidth}
\centerline{\includegraphics[width=0.95\linewidth]{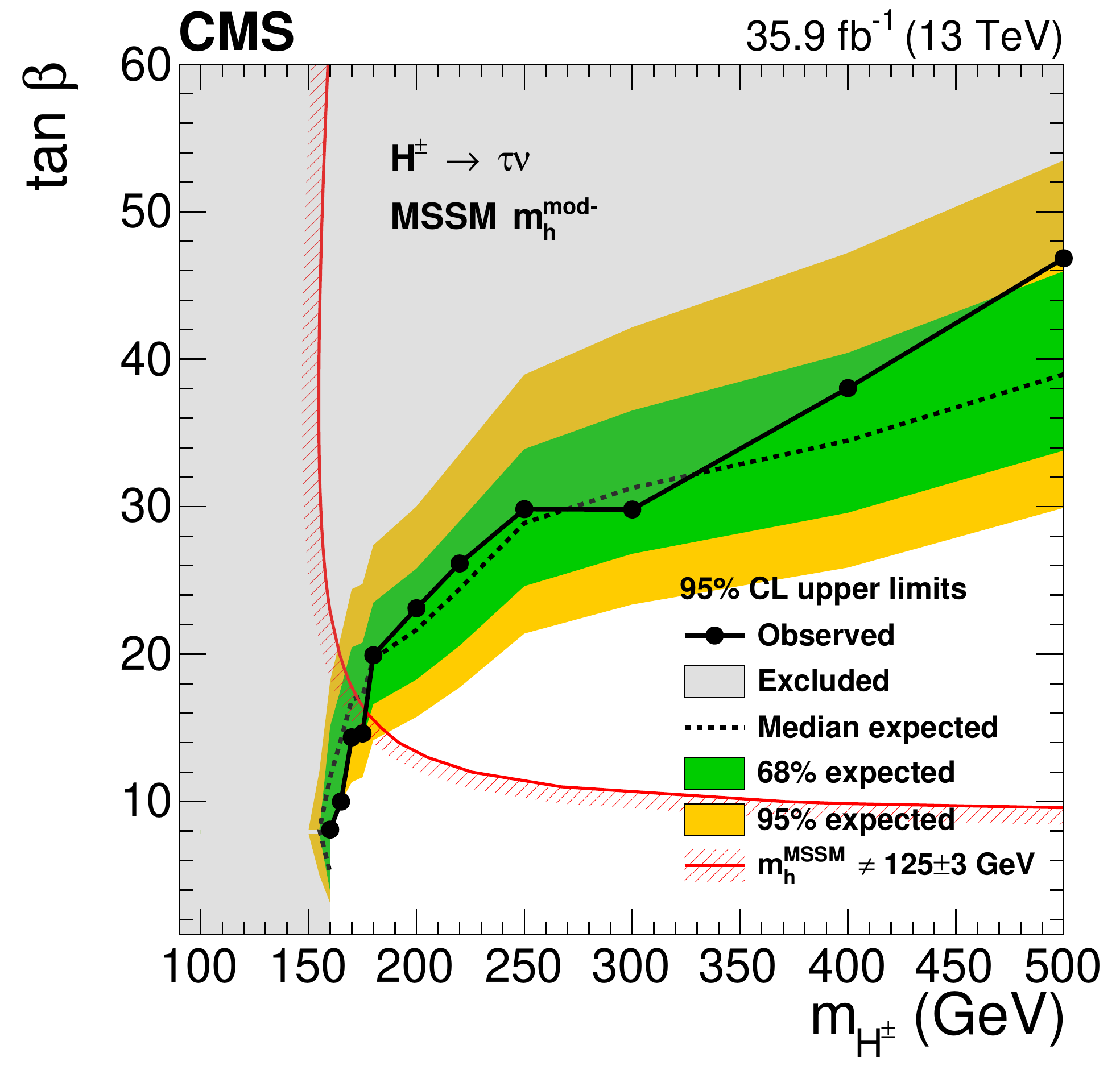}}
\end{minipage}
\hfill
\begin{minipage}{0.32\linewidth}
\centerline{\includegraphics[width=0.95\linewidth]{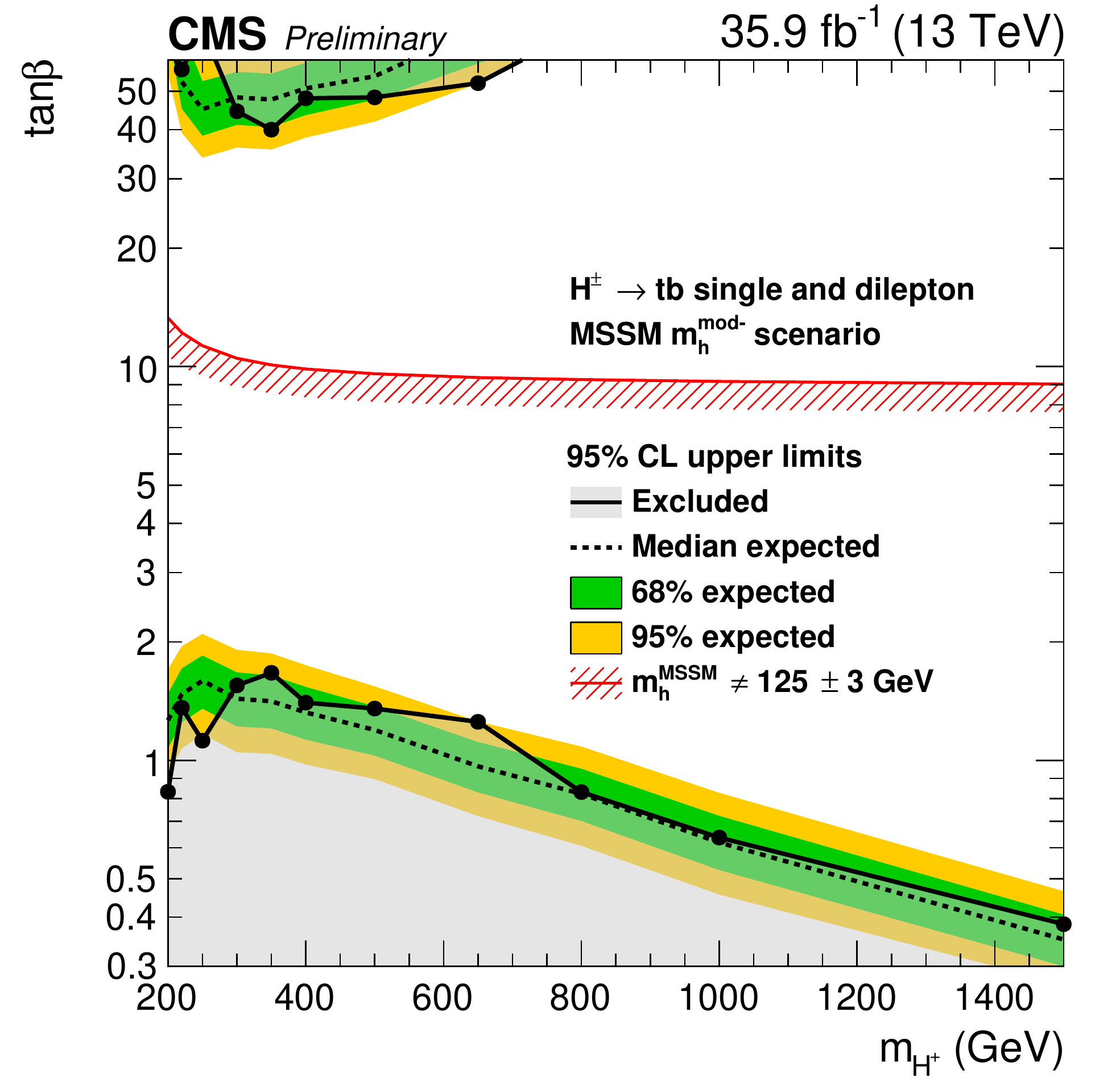}}
\end{minipage}
\hfill
\begin{minipage}{0.32\linewidth}
\centerline{\includegraphics[width=0.95\linewidth]{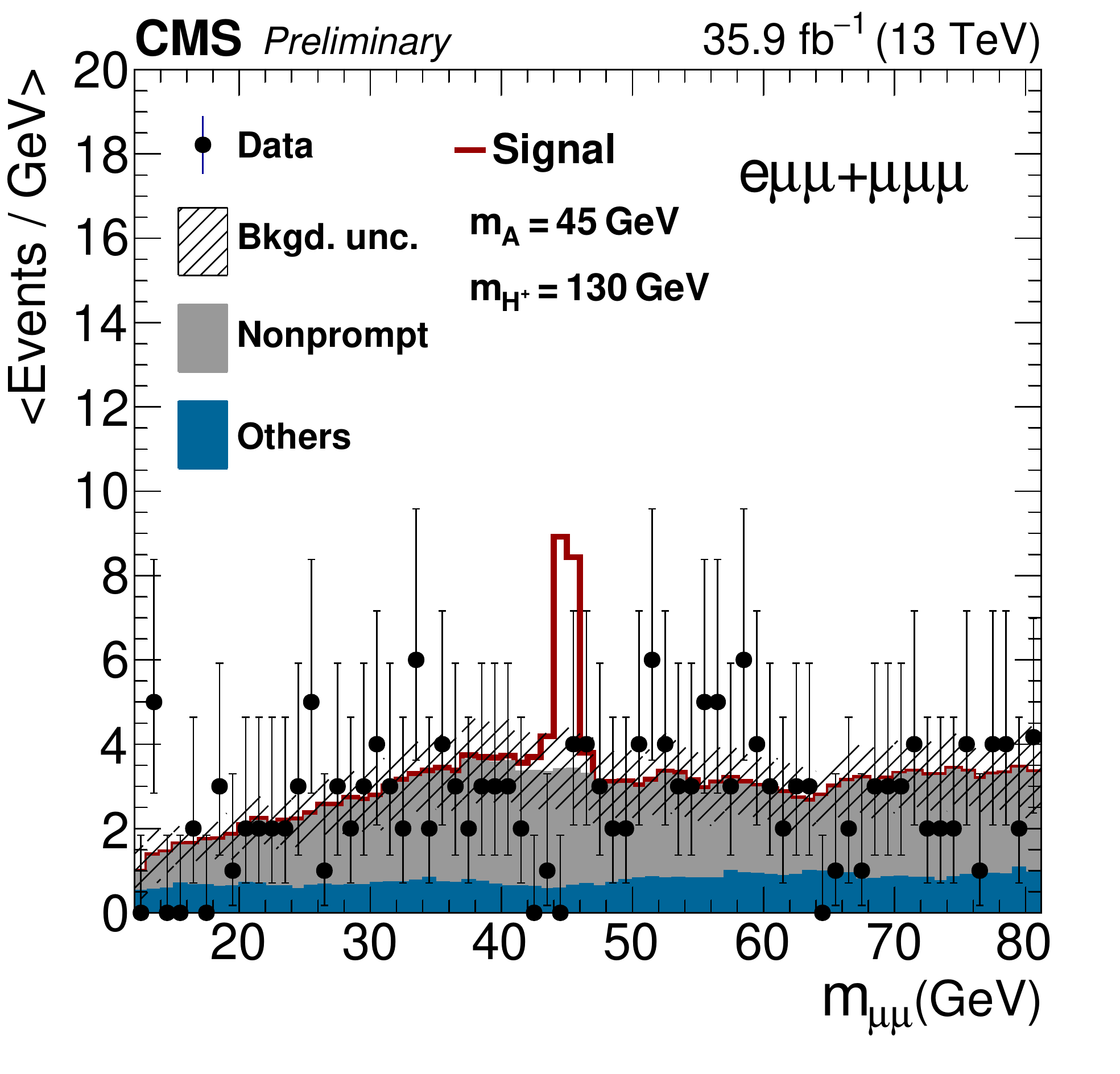}}
\end{minipage}
\caption[]{For the MSSM $m_\mathrm{h}^\mathrm{mod-}$ scenario, the $\mathrm{H}^+ \to \tau\nu$ exclusion region\cite{hptaunu} (left) and 
the $\mathrm{H}^+ \to$tb exclusion region\cite{hptb} (center). Dimuon mass distribution for the $\mathrm{H}^+ \to AW$ analysis\cite{hpaw} (right).
}
\label{fig3}
\end{figure}

Decays to $\tau\nu$ are the flagship LHC channel for the search for charged Higgs bosons, showing the highest sensitivity in the mid- and high-$\tan\beta$ regions. 
Three channels are combined: $\tauhad$+jets, $\tauhad + 1\ell$ and $\tauhad+0\ell$ and the analysis covers the low-mass region (where t$\to$bH$^+$ dominates), the 
high-mass region (gb$\to$tH$^+$) and, for the first time in CMS, the intermediate region where none of the two contributions can be neglected\cite{hptaunu}. 
Cross section limits for the charged Higgs boson mass range of 80 GeV -- 3000 GeV  are set and also interpreted in the context of the MSSM, see Figure~\ref{fig3}.

\subsection{Charged Higgs boson decays to tb}
Complementary to $\tau\nu$ decays, studying $\mathrm{H}^+ \to$tb is mainly motivated by MSSM scenarios with low $\tan\beta$. The analysis\cite{hptb} 
requires one or two leptons and then categorizes events by the number of jets and b-tagged jets. Machine-learning algorithms help to 
cover the $\mathrm{H}^+$ mass range from 200 GeV to 3000 GeV. The result interpreted in the context of the MSSM is shown in Fig.~\ref{fig3}.

\subsection{Charged Higgs boson decays to AW}
If $\tan\beta$ is low but $\mathrm{H}^+$ are light then $\mathrm{H}^+ \to$AW may dominate with branching ratios close to unity. In this 
study\cite{hpaw} final states with three muons or two muons and an electron are investigated, aiming at A$\to \mu\mu$ decays. The $m_{\mu\mu}$ distribution 
is shown in Fig.~\ref{fig3}. Limits on the branching ratio t$\to$bH$^+$ from 0.6\% to 2.9\% are set for A boson masses 
from 15 GeV to 75 GeV.

\section{Summary}
A wealth of CMS Higgs physics results has been made public in the month prior to the workshop, tackling the SM Higgs hypothesis from two 
directions: precision measurements and searches for additional Higgs bosons. All results are compatible with the SM but watch out for the
full LHC Run 2 results expected this year and the LHC Run 3 to follow in 2021 which will lead to sensitivities allowing to cut deeply into the parameter 
space of many popular BSM scenarios.

\end{document}